\pdfoutput=1
\documentclass[a4paper]{PoS}

\usepackage{amsmath}
\usepackage{amssymb}
\usepackage{mathbbol}
\usepackage{slashed}

\usepackage{picins}
\usepackage{overpic}
\usepackage{url}

\usepackage[format=hang,justification=justified]{caption,subfig}
\newcommand{\units}[1]{\ensuremath{\,\mathrm{#1}}}
\newcommand{\bra}[1]{\left\langle #1 \right|}
\newcommand{\ket}[1]{\left| #1 \right\rangle}

\newcommand{\elll}{\ell}
\newcommand{\prp}{\perp}

\newcommand{\lat}{{\text{sW}}}
\newcommand{\tcdot}{\ensuremath{\!\cdot\!}}

\newcommand{\mquad}{\hspace{2mm}}

\newcommand{\crossout}[1]{\setbox0\hbox{#1} \hbox to \wd0{\rlap{\rule[.6ex]{\wd0}{.5pt}}\box0}}

\newcommand{\Wline}[2]{\ensuremath{{\mathcal{U}}_{[#1,#2]}}}
\newcommand{\WlineC}[1]{\ensuremath{{\mathcal{U}}_{#1}}}

\newcommand{\latop}[1]{\mathcal{O}^{#1}}

\newcommand{\myRe}{\ensuremath{\,\mathrm{Re}\ }}
\newcommand{\myIm}{\ensuremath{\,\mathrm{Im}\ }}

\title{Transverse Momentum Distributions of Quarks in the Nucleon from Lattice QCD }

\ShortTitle{Transverse Momentum Distributions from Lattice QCD}

\author{%
	\speaker{Bernhard~U.~Musch}$^a$,
	Philipp~H\"agler$^a$,
	Andreas~Sch\"afer$^b$,
	Dru~B.~Renner$^c$,
	John~W.~Negele$^d$,
	LHPC (Lattice Hadron Physics Collaboration)\\
	\llap{$^a$}	Institut f\"ur Theoretische Physik T39, Physik-Department, Technische Universit\"{a}t M\"{u}nchen, \\
	James-Franck-Stra{\ss}e, D-85747 Garching, Germany\\
	\llap{$^b$}Institut f\"ur Theoretische Physik, Universit\"at Regensburg, \\
	D-93040 Regensburg, Germany\\
	\llap{$^c$}Theory Group, Deutsches Elektronen-Synchrotron DESY,\\
	Platanenallee 6, D-15738 Zeuthen, Germany \\
	\llap{$^d$}Center for Theoretical Physics, Massachusetts Institute of Technology, \\
		Cambridge, MA02139, USA\\
	E-mail: \email{bmusch@ph.tum.de}%
	}

\abstract{Transverse momentum dependent parton distribution functions (TMDPDFs) encode information about the intrinsic motion of quarks inside the nucleon. They are important non-perturbative ingredients in our understanding of, e.g., azimuthal asymmetries and other qualitative features in semi-inclusive deep inelastic scattering experiments. We present first calculations on the lattice, based on MILC gauge configurations and propagators from LHPC. They yield polarized and unpolarized transverse momentum dependent quark densities and enable us to test the assumption of factorization in x and $k_\prp$. The operators we employ are non-local and contain a Wilson line, whose renormalization requires the removal of a divergence linear in the cutoff $a^{-1}$. }

\FullConference{LIGHT CONE 2008 Relativistic Nuclear and Particle Physics\\
  July 7-11, 2008\\
  Mulhouse, France}

\begin{document}

\section{Introduction}

\begin{figure}[hb]
	\centering
	\begin{minipage}[t]{0.35\textwidth}
	 	\centering
		\includegraphics*[width=4cm,trim=150 0 70 250,clip=true]{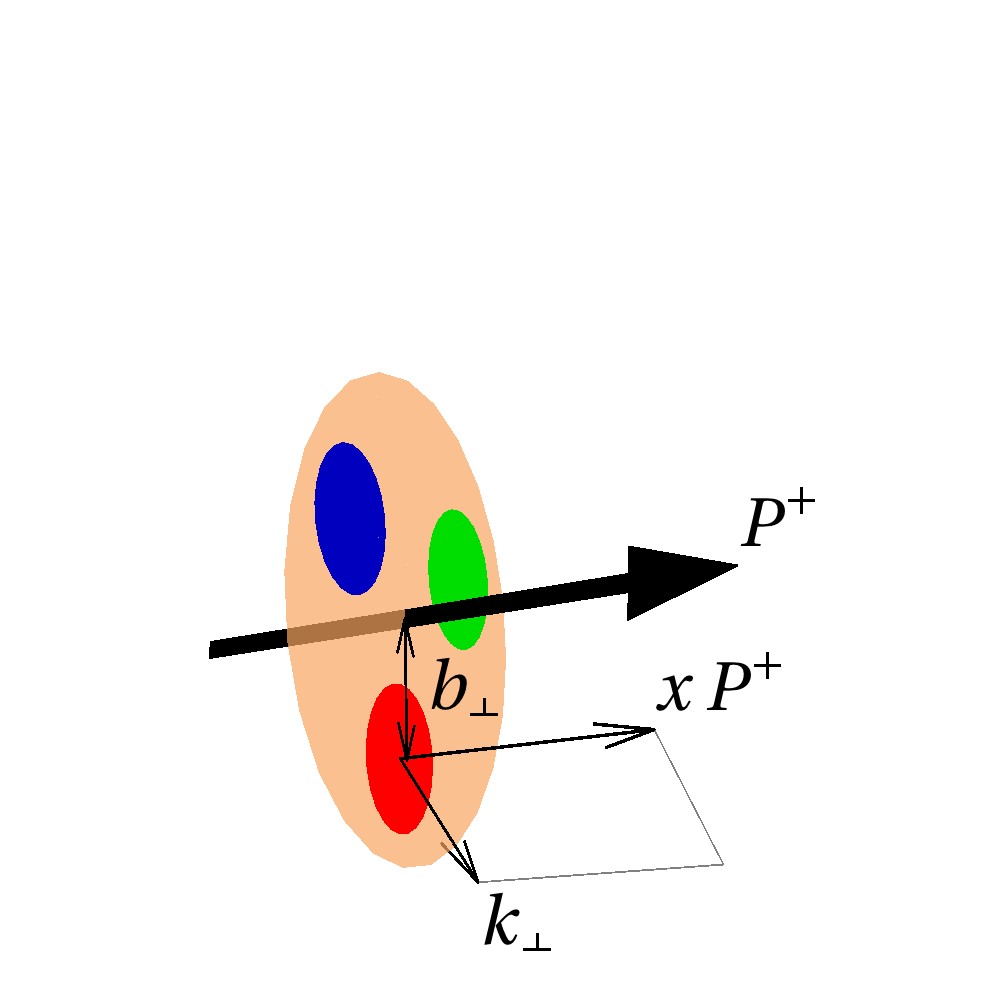}\par
		\caption{}%
		\label{fig-partonpicture}%
		\end{minipage}
	\hfill
	\begin{minipage}[t]{0.63\textwidth}
	 	\centering
		\includegraphics*[width=7.5cm]{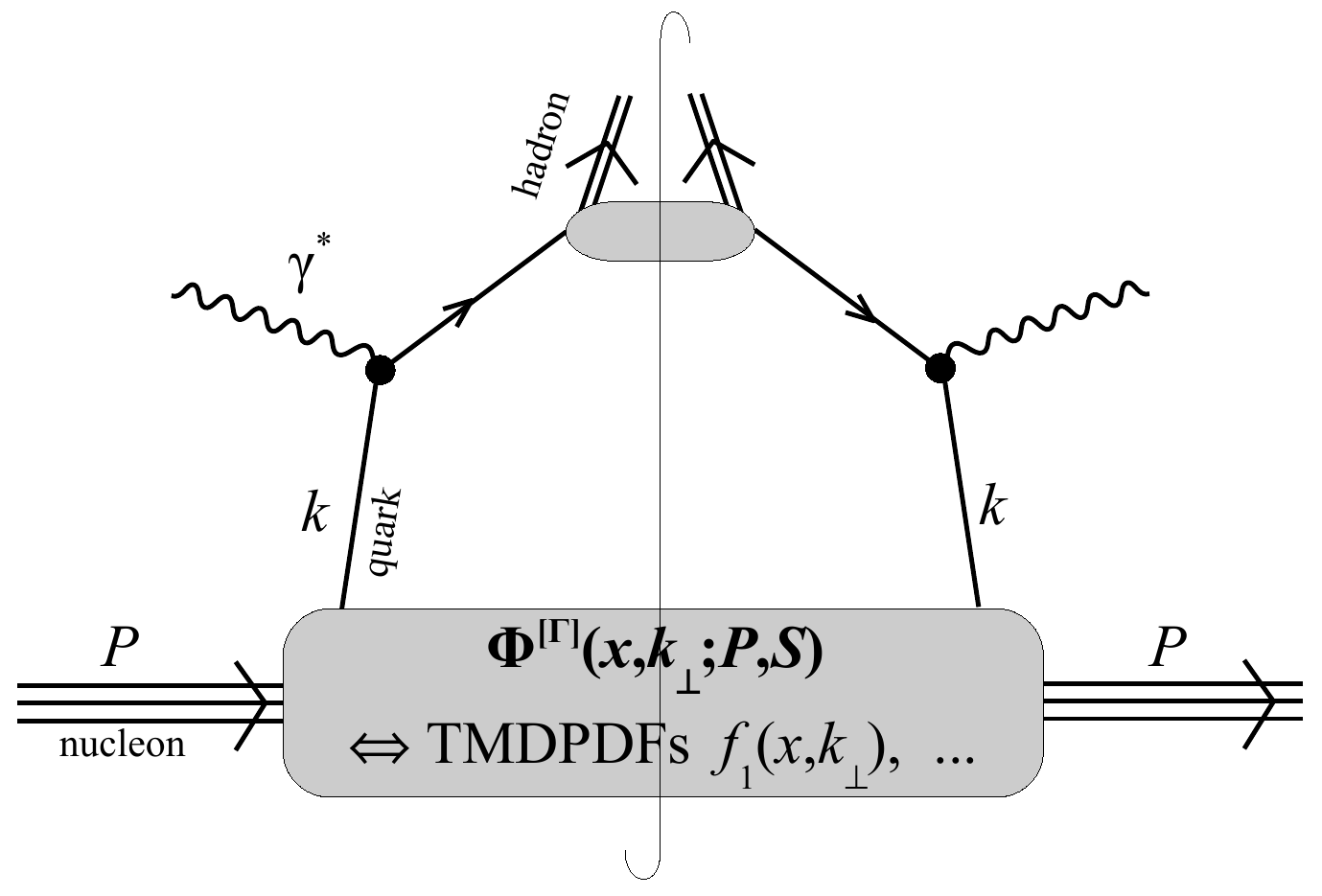}\par
		\caption[SIDIS diagram and gauge links]{%
			Factorized tree level diagram of SIDIS.\par%
			\label{fig-SIDISdiagram}%
			}%
		\end{minipage}
\end{figure}

Semi-inclusive deeply inelastic scattering (SIDIS) experiments are sensitive to many correlations between the direction of parton and hadron spins and intrinsic transverse momenta. These have been parameterized in a systematic manner using transverse momentum dependent parton distribution functions (TMDPDFs), see \cite{Muld95}. TMDPDFs describe the distribution of partons carrying a longitudinal momentum fraction $x$ and an intrinsic transverse momentum $k_\prp$ in a hadron as illustrated in Fig.~\ref{fig-partonpicture}. Here we give an update of our effort \cite{Musch:2007ya} to develop techniques suitable for the calculation of moments of TMDPDFs on the lattice. Note that TMDPDFs are not to be confused with generalized parton distribution functions (GPDs), which provide probability distributions with respect to the impact parameter $b_\prp$ rather than $k_\prp$. For an overview of recent hadron structure studies in lattice QCD, we refer to \cite{Zanotti}.

Fig.~\ref{fig-SIDISdiagram} illustrates the factorization of SIDIS into perturbative and non-perturbative parts. The lower blob represents the non-perturbative contribution of the nucleon and is described by
\begin{equation}
	\Phi^{[\Gamma]}(x,k_\prp;P,S) \equiv 
	\frac{1}{2} \int dk^- \int \frac{d^4\elll}{(2\pi)^4}\ 
	e^{-ik \cdot \elll}\ 
	\bra{P,S}\ \bar q(\elll)\, \Gamma\ \WlineC{\mathcal{C}(\elll,0)}\ q(0)\ \ket{P,S}%
	, 
	\label{eq-corr}
\end{equation}
where $\ket{P,S}$ represents a nucleon state of momentum $P$ and spin $S$, $\Gamma$ is a Dirac matrix and $k$ is the quark momentum, with $k^+=xP^+$. The Wilson line $\WlineC{\mathcal{C}(\elll,0)}$ connecting the quark operators ensures gauge invariance. In SIDIS, $\WlineC{\mathcal{C}(\elll,0)}=\Wline{\elll}{\elll+\infty \hat n_-}\Wline{\elll+\infty \hat n_-}{\infty \hat n_-}\Wline{\infty \hat n_-}{0}$ is a concatenation of three straight Wilson lines running to light cone infinity and back \cite{Belitsky:2002sm,Boer:2003cm}.

Some examples of TMDPDFs are $f_1$, $g_{1L}$ and $g_{1T}$, defined via \cite{Muld95}
\begin{align}
	\Phi^{[\gamma^+]}(x, k_\prp;P,S) &\ =\ f_1(x,k_\prp^2) \label{eq-phivect} \\
	\Phi^{[\gamma^+ \gamma^5]}(x, k_\prp;P,S) &\ =\ \frac{m_N}{P^+}S^+\, g_{1L}(x,k_\prp^2)\ +\ \frac{k_\prp \cdot S_\prp}{m_N}\ g_{1T}(x,k_\prp^2)%
	.
	\label{eq-phiaxialvect}
\end{align}

\section{TMDPDFs from the Euclidean lattice}
In this exploratory study, we simplify the calculation and employ a single straight Wilson line $\WlineC{\mathcal{C}(\elll,0)}=\Wline{\elll}{0}$ running from $0$ to $\elll$. In this case, the matrix element appearing in eq.\,(\ref{eq-corr}) can be evaluated directly on the Euclidean lattice, as long as we set $\elll^0 = 0$. To facilitate the translation of the correlators evaluated on the lattice into TMDPDFs, we first parameterize the matrix element in terms of Lorentz-invariant amplitudes $\tilde A_i(\elll^2,\elll \cdot P)$, for example
\begin{align}
	\bra{P,S}\ \overline{q}(\elll)\, \gamma_\mu\, \mathcal{U}\, q(0)\ \ket{P,S}
		& =
		4\ \tilde{A}_2\ P_\mu
		+ 4i\,{m_N}^2\ \tilde{A}_3\ \elll_\mu %
	,\label{eq-gammamutrace}\\
	\bra{P,S}\ \overline{q}(\elll)\, \gamma_\mu \gamma^5\, \mathcal{U}\, q(0)\ \ket{P,S}
		& = 
		- 4\, m_N\ \tilde{A}_6\ S_\mu
		- 4i\,m_N\ \tilde{A}_7\ P_\mu\, \elll \tcdot S 
		+ 4\,{m_N}^3\ \tilde{A}_8\ \elll_\mu\, \elll \tcdot S%
	. 
	\label{eq-phitildetraces}
	\end{align}
The amplitudes $\tilde A_i(\elll^2,\elll \tcdot P)$ are extracted on the lattice and then Fourier transformed into TMDPDFs.
For example, from eqns.\,(\ref{eq-phivect}), (\ref{eq-corr}) and (\ref{eq-gammamutrace}) we get
\begin{equation}
	f_1(x,k_\prp^2)  = \int \frac{d(\elll \tcdot P)}{2\pi}\ e^{-i\, \elll \tcdot P\, x}  \int_0^\infty \frac{d(-\elll^2)}{4 \pi}\ J_0(\sqrt{-\elll^2}\, |k_\prp|)\ 2\,\tilde{A}_2(\elll^2,\elll\tcdot P)%
	,\label{eq-f1ft}
\end{equation}
where $J_0$ is a Bessel-function. The restriction to $\elll^0=0$ on the lattice translates into the constraints
\begin{equation}
	\elll^2 \leq 0, \qquad |\elll \cdot P| \leq |\vec \elll| |\vec P|,
	\label{eq-latconstraints}
\end{equation}
which preclude us from evaluating the full $x$- and $k_\prp$-dependence directly as in eq.\,(\ref{eq-f1ft}), but are harmless if we are only interested in the first Mellin moment, i.e., if we integrate over x. For example, we obtain 
\begin{align}
	f_1^{(1)}(k_\prp^2) & \equiv \int_{-1}^{1} dx\ f_1(x,k_\prp^2) = \int_0^\infty \frac{d(-\elll^2)}{4\pi}\  
	J_0(\sqrt{-\elll^2}\,|k_\prp|)\ 2 \tilde{A}_2(\elll^2,\elll \tcdot P = 0), \label{eq-f1} \\
	g_{1T}^{(1)}(k_\prp^2) & \equiv \int_{-1}^{1} dx\ g_{1T}(x,k_\prp^2) = \int_0^\infty \frac{d(-\elll^2)}{4\pi} \frac{J_1(\sqrt{-\elll^2}\,|k_\prp|)}{\sqrt{-\elll^2}\,|k_\prp|} \elll^2 m_N^2\ 2 \tilde{A}_7(\elll^2,\elll \tcdot P = 0)%
	.
	\label{eq-g1T}
\end{align}

\section{Simulation technique and parameters}

\begin{floatingfigure}
	\includegraphics[width=.25\textwidth,clip=true,trim=0 10 0 10]{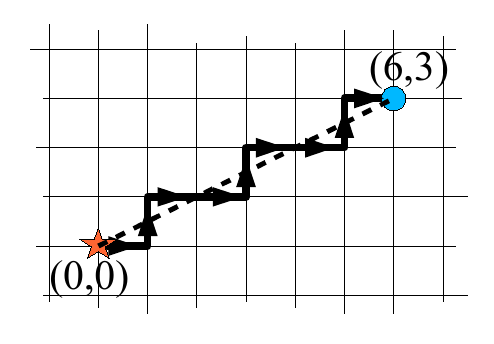}\par
	\vspace{-3.7mm}
	\caption{}
	\label{fig-steplike}
\end{floatingfigure}
The matrix element $\bra{P,S}\, \bar{q}(\elll)\, \Gamma\, \Wline{\elll}{0}\, q(0)\, \ket{P,S}$ is evaluated using ratios of three- and two-point functions as described in Ref.~\cite{Musch:2007ya}. 
The non-local operator $\latop{\Gamma}(\elll) \equiv \bar{q}(\elll)\, \Gamma\, \Wline{\elll}{0}\, q(0)$ inserted in the three-point function 
contains the Wilson-line $\Wline{\elll}{0}$%
, which is implemented as a product of link variables. For oblique angles, we approximate a straight line by a step-like path as illustrated in Fig.~\ref{fig-steplike}.

For our studies we work with MILC gauge configurations \cite{Ber01} based on an AsqTad improved staggered quark action with 2+1 flavors on a $20^3 \times 64$ lattice with a lattice spacing $a \approx 0.12 \units{fm}$ and a strange quark mass $a m_s = 0.050$. We have used three different light quark masses, $a m_{u,d} = 0.020$ ($m_\pi \approx 500 \units{MeV}$, $239$ configurations), $a m_{u,d} = 0.030$ ($m_\pi \approx 600 \units{MeV}$, $281$ configurations) and the three-flavor degenerate case $a m_{u,d} = 0.050$ ($m_\pi \approx 760 \units{MeV}$, $213$ configurations). 

The gauge configurations have been HYP smeared and bisected in the temporal direction to double statistics.
We are using domain wall propagators and sequential propagators previously calculated by the LHPC collaboration on these configurations, with the valence quark mass tuned to match the staggered sea (see, e.g., \cite{Hagler:2007xi}). The sequential propagators feature a source-sink separation of $t_\text{sink} - t_\text{source} = 10$, and are available for two lattice nucleon momenta $\vec P=(0,0,0)$ and $\vec P=(-1,0,0)$, the latter corresponding to $500\units{MeV}$ in physical units. We neglect contributions from disconnected diagrams. We have developed our software using the Chroma library \cite{Edwards:2004sx}.

\section{Renormalization of the Wilson Lines}
\label{sec-renorm}

The Wilson line $\Wline{\elll}{0}$ in our non-local operator gives rise to a linear divergence, which has to be removed by a renormalization constant $\delta m$ proportional to the cutoff, given by $a^{-1}$ on the lattice. Refs. \cite{Craigie:1980qs,Dorn:1986dt} show within continuum theory that the renormalized operator is of the form
\begin{equation}
	\latop{\Gamma}_\text{ren}(\elll) = 
	Z^{-1} \exp(-\delta m \ L)\ \ \latop{\Gamma}(\elll)%
	.
	\label{eq-Wlineren}
\end{equation}
Here $Z^{-1}$ subsumes renormalization factors associated with divergences at the end points, and $L$ is the total length of the smooth Wilson line. 

In lattice QCD, the linear divergence has been a long standing issue in the context of heavy quark propagators \cite{Maiani:1991az}. We have calculated $a \delta m$ for link paths on the axes in leading order perturbation theory, adapting the procedure in Refs. \cite{Eichten:1989kb,Boucaud:1989ga} to our action according to Refs. \cite{Bistro,DeGrand:2002va}. 

However, perturbation theory is not expected to give accurate results. Therefore we have also sought to determine $a \delta m$ non-perturbatively with our ``taxi driver method'', which is based on the assumption that on the lattice, $L$ in eq.~(\ref{eq-Wlineren}) is given by the total number of link variables, and thus allows us to deduced the renormalization constants from the comparison of straight and step-like link paths. There are two variants of this method, one based on data from Wilson lines in a Landau gauge fixed ensemble, and one based on Wilson loops. For the moment, we assume that the quark mass dependence is weak, and use the constants determined on the three-flavor degenerate lattice also for the lighter quark masses. We are currently testing the validity of the taxi driver approach and alternative methods on several different lattice spacings.

\section{Results with Preliminary Renormalization}

\begin{figure}[bt]
	\centering
	\subfloat[][]{%
		\label{fig-A2}%
		\begin{overpic}[width=0.49\textwidth,trim=0 0 10 0,clip=true]{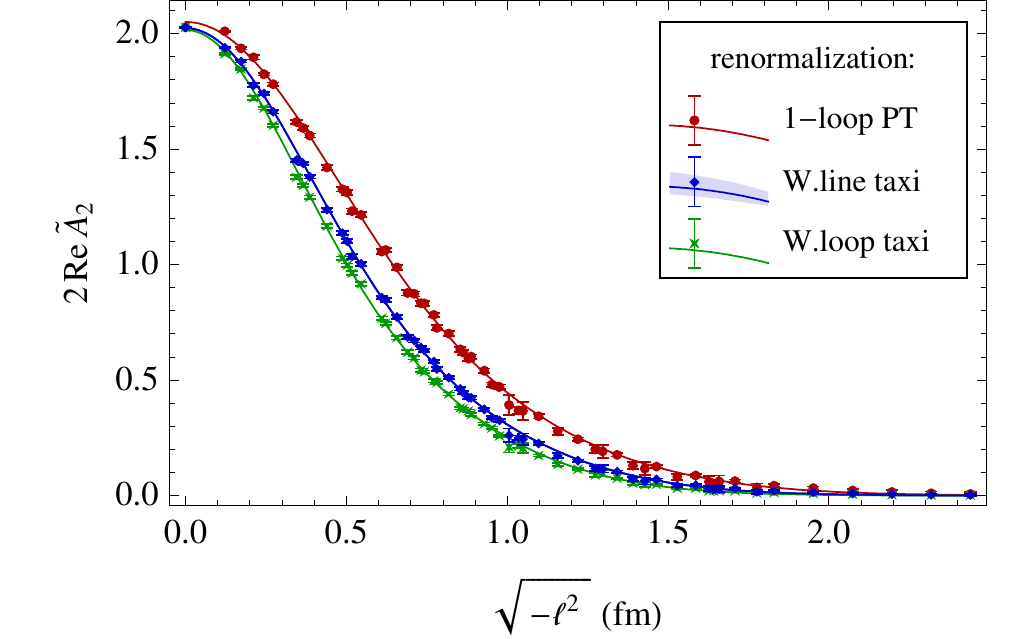}
			\put(32,55){\begin{minipage}{3cm}{\small up quarks\\$m_\pi=500\units{MeV}$\par}\end{minipage}}
			\end{overpic}}%
	\hfill%
	\subfloat[][]{%
		\label{fig-A7}%
		\begin{overpic}[width=0.5\textwidth,trim=0 0 10 0,clip=true]{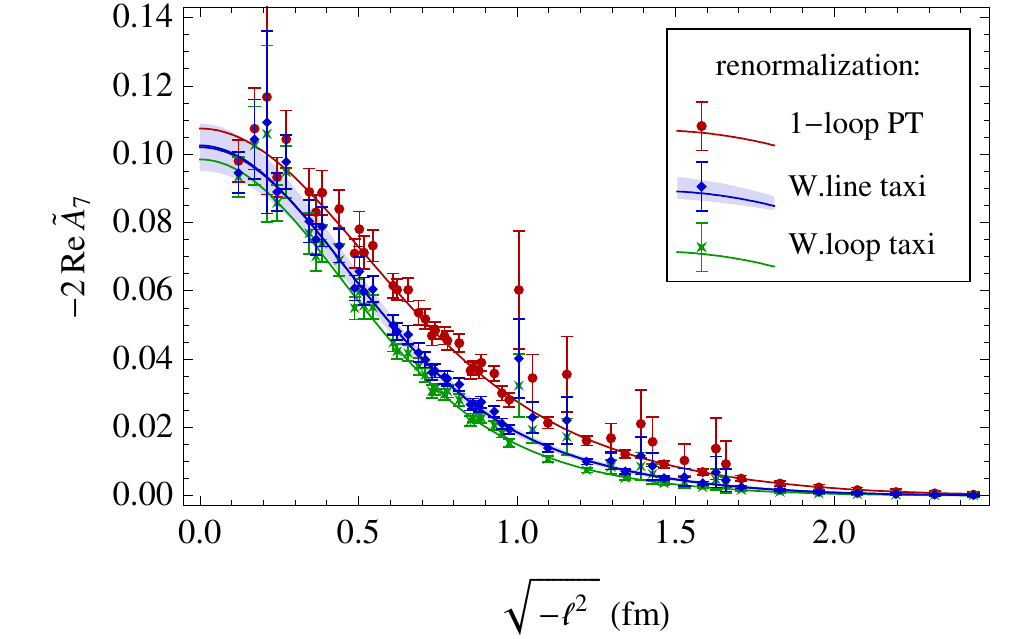}
			\put(32,55){\begin{minipage}{3cm}{\small up quarks\\$m_\pi=500\units{MeV}$\par}\end{minipage}}
			\end{overpic}}
	\caption[unpolarized]{%
		\subref{fig-A2}\mquad%
			Amplitude $2 \myRe \tilde{A}_2( \elll^2, \elll\tcdot P = 0)$ for up quarks. The continuous curves are fits of the form $C_1 \exp(-\elll^2/\sigma_1^2) + C_2 \exp(-\elll^2/\sigma_2^2)$. We have used three different sets of renormalization constants: For the upper curve, we have determined $a \delta m$ from 1-loop perturbation theory, for the two lower curves we have employed two variants of the taxi driver method.
			\par%
		\subref{fig-A7}\mquad%
			The same for the amplitude $-2 \myRe \tilde{A}_7( \elll^2, \elll\tcdot P = 0)$.
			\par%
		\label{fig-amps}
		}
	\end{figure}

Here we present some results obtained by applying the techniques sketched above. Note that at the present stage we regard our renormalization procedure still as preliminary. In the following, we label our distributions ``sW'' to indicate that they are based on straight Wilson lines and are therefore not strictly identical to the TMDPDFs defined and used in the literature and for the description of, e.g., SIDIS.

\begin{figure}[b]
	\subfloat[][]{%
		\label{fig-f1}%
		\begin{overpic}[width=0.32\textwidth,trim=0 0 0 0,clip=true]{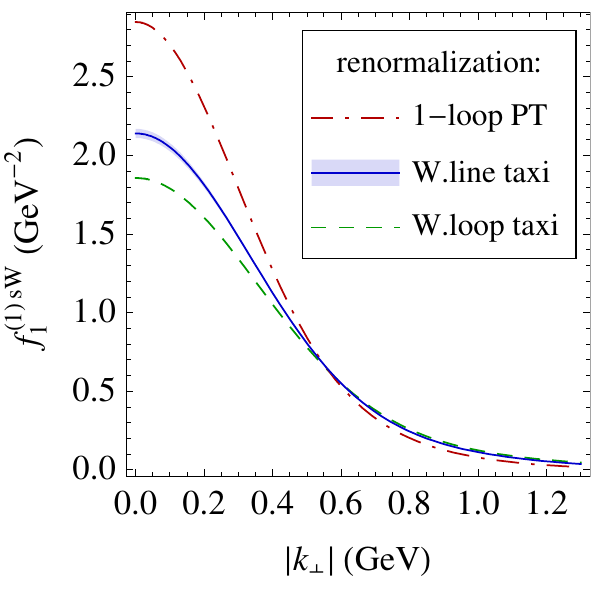}
			\put(35,42){\begin{minipage}{3cm}{\small\flushright
			$m_\pi=500\units{MeV}$\\%
			up quarks\\
			\par}\end{minipage}}
			\end{overpic}\par
		}
	\hfill
	\subfloat[][]{%
		\label{fig-g1T}%
		\begin{overpic}[width=0.32\textwidth,trim=0 0 0 0,clip=true]{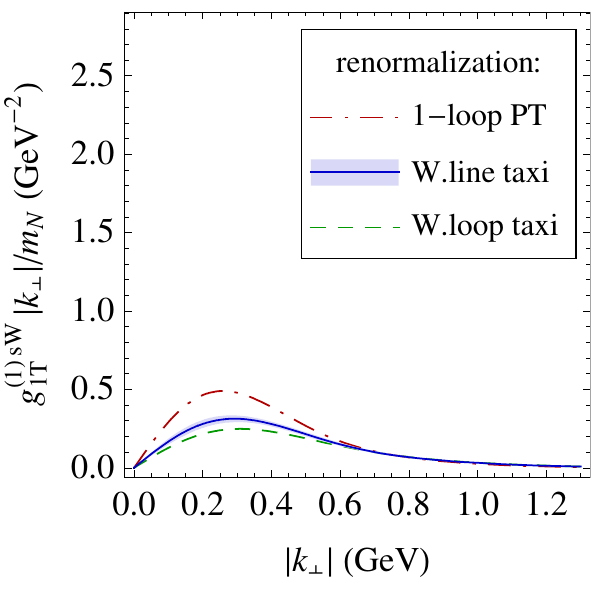}
			\put(35,42){\begin{minipage}{3cm}{\small\flushright
			$m_\pi=500\units{MeV}$\\%
			up quarks\\
			\par}\end{minipage}}
			\end{overpic}\par
		}
	\hfill
	\subfloat[][]{%
		\label{fig-rmskt}%
		\includegraphics[width=0.32\textwidth]{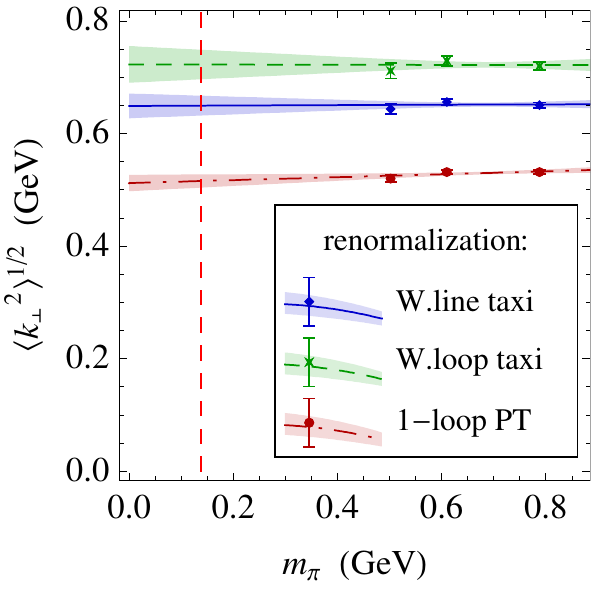}
		}
	\caption[unpolarized]{%
		\subref{fig-f1}\mquad%
			$f^{(1)\text{sW}}_1(k_\prp^2)$ for up quarks as obtained from the Fourier transform of the fits in Fig.~\ref{fig-A2}. 
			\par%
		\subref{fig-g1T}\mquad%
			$\frac{|k_\prp|}{m_N}\, g^{(1)\text{sW}}_{1T}(k_\prp^2)$ for up quarks as obtained from the fits in Fig.~\ref{fig-A7}. 
			\par%
		\subref{fig-rmskt}\mquad%
			Linear chiral extrapolation of the root mean squared transverse momentum $\langle k_\prp^2 \rangle^{1/2}$ from $f^{(1)\text{sW}}_1(k_\prp^2)$ for up quarks minus down quarks.\par
		\label{fig-f1related}
		}
	\end{figure}
\begin{figure}[hbt]
	\centering
	\begin{overpic}[width=0.85\textwidth,trim=10 20 20 20,clip=true]{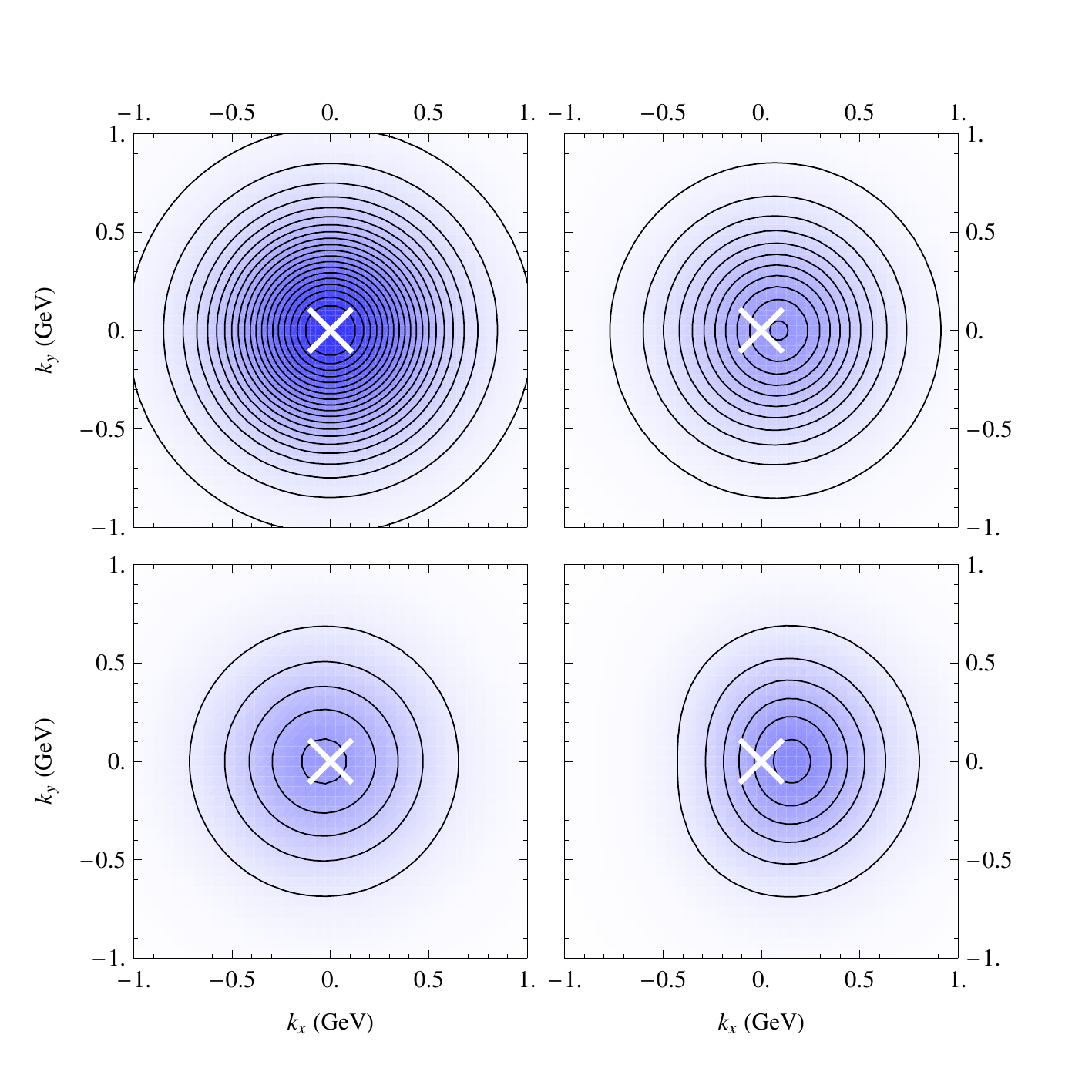}
		\put(12.5,51){\subfloat[][]{\label{fig-unpolDens}}}
		\put(44.5,86){up}
		\put(55,51){\subfloat[][]{\label{fig-polDensU}}}
		\put(87,86){up}
		\put(54.5,80.5){\includegraphics[width=1.6cm]{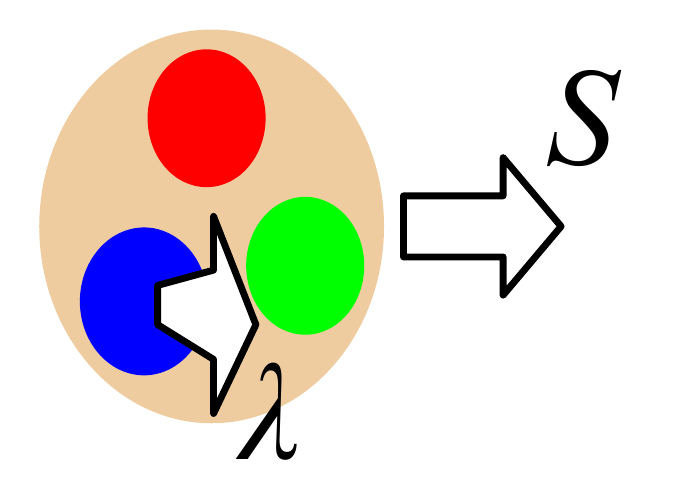}}
		\put(12.5,8.5){\subfloat[][]{\label{fig-polDensD}}}
		\put(41,42.5){down}
		\put(11.5,37.5){\includegraphics[width=1.6cm]{figs/PolarizedNucleon2.pdf}}
		\put(55,8.5){\subfloat[][]{\label{fig-polDensUminusD}}}
		\put(86,42.5){u-d}
		\put(54.5,37.5){\includegraphics[width=1.6cm]{figs/PolarizedNucleon2.pdf}}
		\end{overpic}
	\caption[polarized]{%
		Quark density plots. Here the renormalization constants have been chosen according to the taxi driver method based on Wilson lines.\par
		\subref{fig-unpolDens}\mquad%
			$f^{(1)\text{sW}}_1(k_\prp^2) = \int dx\, \Phi^{[\gamma^+]}(x,k_\prp)$ for up quarks at $m_\pi = 500\units{MeV}$. 
			We interpret this as the charge density of up quarks in the nucleon in the transverse momentum plane. \par
		\subref{fig-polDensU}\mquad%
			Density $\rho(k_\prp,S_\prp)$ of up quarks with positive helicity $\lambda\!=\!+1$ (i.e., with spin pointing in $z$-direc\-tion) in a nucleon polarized in transverse x-direction $S_\prp = (1,0)$, evaluated at a pion mass $m_\pi = 500\units{MeV}$. The distribution features an average transverse momentum shift $\langle k_x \rangle = (67 \pm 5_\text{stat} \pm 3_\text{renorm.})\units{MeV}$, where the uncertainty from renormalization has been estimated from the comparison of the three different sets of renormalization constants employed in Figs.~\ref{fig-amps} and \ref{fig-f1related}.
			\par%
		\subref{fig-polDensD}
			Same as in \subref{fig-polDensU} but for down quarks. The average transverse momentum shift $\langle k_x \rangle = (-24 \pm 5_\text{stat} \pm 3_\text{renorm.})\units{MeV}$ has the opposite sign as for up quarks. \par
		\subref{fig-polDensUminusD}\mquad%
			Same as in \subref{fig-polDensU} for up quarks minus down quarks. The deformation appears amplified. Note that this is not a density and not necessarily positive. 
			\par%
		\label{fig-dens}
		}
	\end{figure}

In Fig.~\ref{fig-A2} we display results for $2 \tilde A_2(\elll^2, \elll \tcdot P=0)$. They are obtained from a three-point function with the operator $\latop{\gamma^4}(\elll)$, where the Wilson line is renormalized using different approaches as described above. The overall normalization can be obtained requiring charge conservation, $2 \tilde A^{u-d}_2(0,0)=1$. Applying eq.\,(\ref{eq-f1}) to the fit results, we get $f_1^{(1)\text{sW}}(k_\prp^2)$ as plotted in Fig.~\ref{fig-f1}, which is interpreted as the unpolarized distribution of quarks in the unpolarized nucleon in Fig.~\ref{fig-unpolDens}. In the unpolarized channels, the nucleon looks axially symmetric. For the large pion masses currently analyzed, the quark mass dependence of the width of this distribution appears to be rather weak, see Fig.~\ref{fig-rmskt}. 

The axial symmetry is distorted in the polarized case. As an example, we can consider the distribution of quarks with positive helicity $\lambda\!\!=\!\!+1$, corresponding to an operator $\latop{\Gamma}(\elll)$ with $\Gamma = \gamma^+ \frac{1}{2}(1+\gamma^5)$ in a nucleon transversely polarized with transverse spin $S_\prp$. Using eqns.\,(\ref{eq-phivect}),\,(\ref{eq-phiaxialvect}) 
\begin{equation}
	\rho(k_\prp,S_\prp)\ \equiv\ \int_{-1}^1 dx\ \Phi^{[\gamma^+(1+\gamma^5)/2]}(x, k_\prp;P,S) =\ 
	\frac{1}{2} \left( f^{(1)}_1(k_\prp^2)\ +\ \frac{k_\prp \cdot S_\prp}{m_N}\ g^{(1)}_{1T}(k_\prp^2) \right)%
	.
	\label{eq-poldens}
\end{equation}
This density is axially asymmetric due to the contribution from the TMDPDF $g_{1T}$, which contains information about the correlation of spins and momenta of the form $\vec s \tcdot \vec P\ k \tcdot S \sim \lambda\, k_\prp \tcdot S_\prp$. According to eq.\,(\ref{eq-g1T}), we obtain $g_{1T}^{(1)\text{sW}}(k_\prp^2)$ from the amplitude $\tilde A_7$, as shown in the case of up quarks in Fig.~\ref{fig-A7} and \ref{fig-g1T}. For down quarks, the results are of opposite sign and smaller by a factor of about $1/6$. In terms of the density $\rho(k_\prp,S_\prp)$, we find that quarks of a specific polarization can have a non-vanishing average transverse momentum
\begin{equation}
  \langle k_\prp \rangle = \frac{\int d^2 k_\prp\ k_\prp\, \rho(k_\prp,S_\prp)}{\int d^2 k_\prp\ \rho(k_\prp,S_\prp)}
\end{equation}
and that it is opposite in sign for up- and down quarks, see Figs.~\ref{fig-polDensU} and \ref{fig-polDensD}. That such deformed quark densities are to be expected has been mentioned, e.g., in Ref.~\cite{Miller:2007ae}. Similar deformations have also been observed in the framework of GPDs in \cite{Diehl:2005jf,Gockeler:2006zu}.

\section{Testing factorization in $x$ and $k_\prp$}

\begin{figure}[b]
	\subfloat[][]{%
		\label{fig-ReA2-cuts}%
		\includegraphics[width=0.48\textwidth]{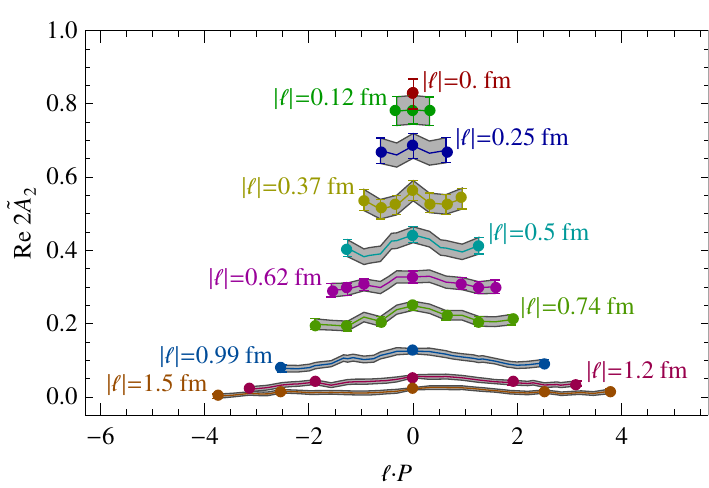}
		}
	\hfill
	\subfloat[][]{%
		\label{fig-ImA2-cuts}%
		\includegraphics[width=0.48\textwidth]{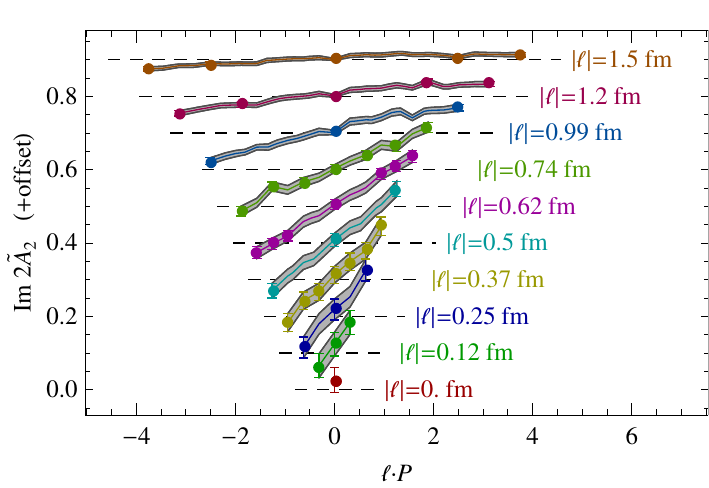}
		}
	\caption[unrenormalized amplitudes]{%
		Cuts through the unrenormalized amplitude $\tilde A_2(\elll^2,\elll\tcdot P)$ at constant $\elll^2$ for $m_\pi \approx 600 \units{MeV}$ and for up minus down quarks. The gray bands are obtained from a linear interpolation on the $(\elll^2,\elll \tcdot P)$-plane. \subref{fig-ReA2-cuts}: Real part. \subref{fig-ImA2-cuts}: Imaginary part. For the sake of clarity, we have added offsets in the ordinate. The dashed lines indicate the respective zero lines. \par%
		\label{fig-A2ldp}
		}
	\end{figure}

\begin{floatingfigure}
	\includegraphics[width=0.5\textwidth,clip=true,trim=12 0 12 0]{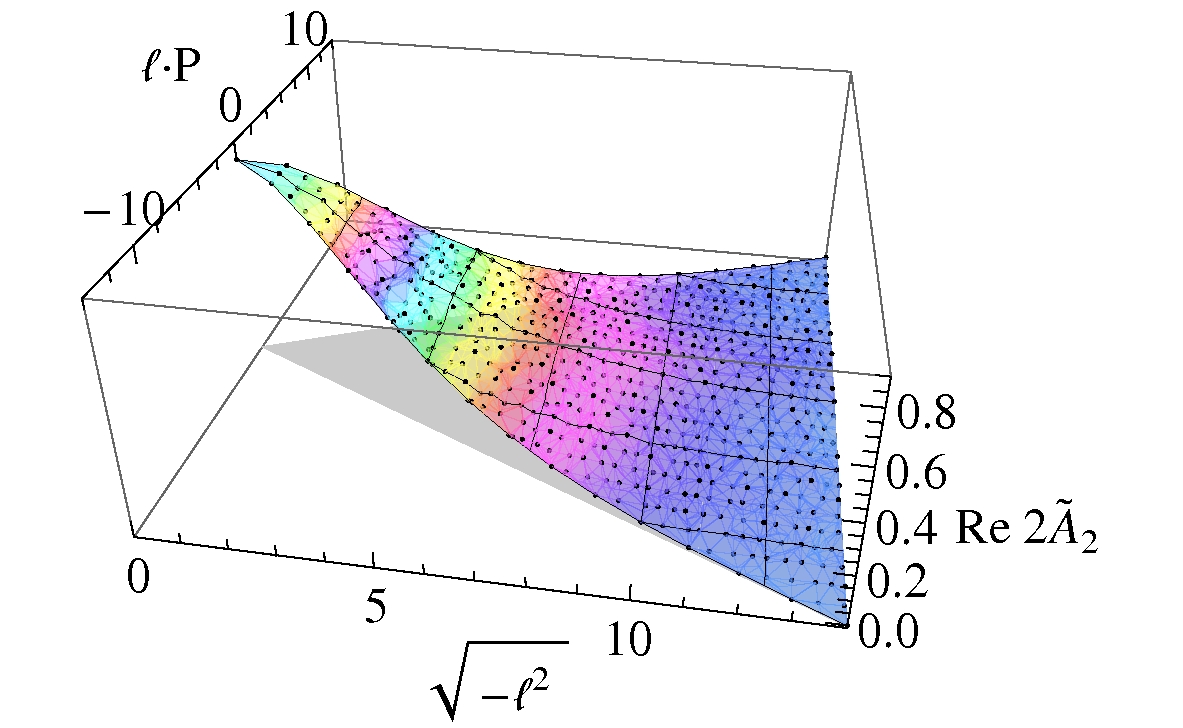}
	\caption{}
	\label{fig-ReA2-3D}%
\end{floatingfigure}
In previous sections we have studied lattice data for $\elll \tcdot P=0$. Let us now explore the $\elll \tcdot P$-dependence of $\tilde A_2(\elll^2,\elll \tcdot P)$. The first integral in eq.\,(\ref{eq-f1ft}) shows that it is related to the Bjorken-$x$-dependence of $f_1^\lat(x,k_\prp)$ via a Fourier transformation.
Figure~\ref{fig-ReA2-3D} gives an overview of the unrenormalized data available for $\myRe \tilde A_2$ for up minus down quarks. The sector with data points is constrained by eq.\,(\ref{eq-latconstraints}) and the largest available nucleon momentum $|\vec P|$ of about $500\units{MeV}$. Figure~\ref{fig-A2ldp} displays the $\elll\tcdot P$-dependence of the real and imaginary parts of $\tilde A_2$ and reveals consistency of our results with $\tilde A_i(\elll^2, \elll \tcdot P) = \tilde A_i^*(\elll,- \elll \tcdot P)$, which follows from the transformation property of the matrix element under Hermitian conjugation.

\newcommand{\fonex}{\ensuremath{\hat{\mathbf{f}}_1^{\,\lat}\!}}
\newcommand{\AtwoldP}{\ensuremath{\hat{\mathbf{A}}_2}}

\begin{figure}[t]
	\begin{minipage}[c]{0.44\textwidth}
		\subfloat[][]{%
			\label{fig-facttest}%
			\includegraphics[width=\textwidth]{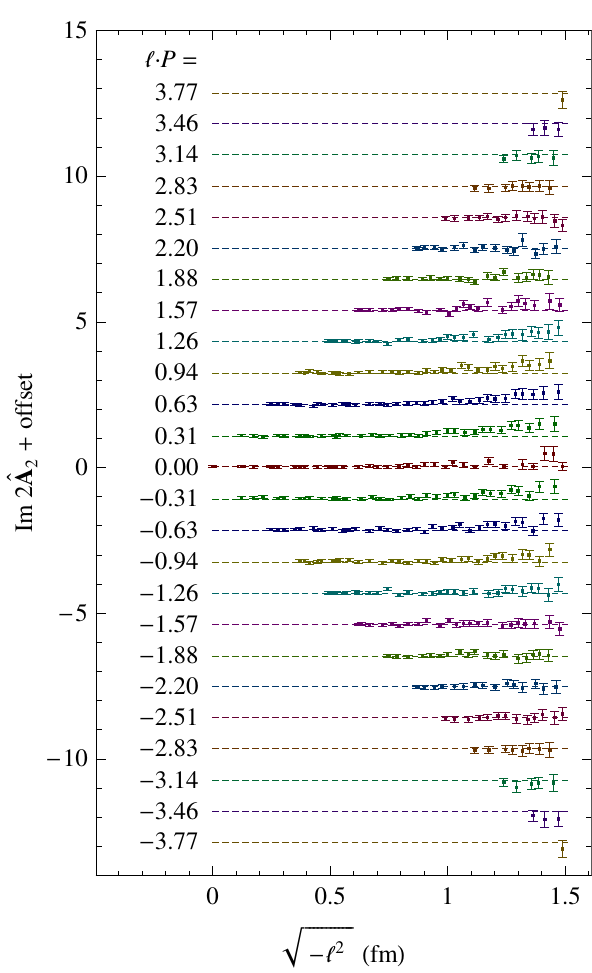}
			}
	\end{minipage}\hfill
	\begin{minipage}[c]{0.54\textwidth}
		\subfloat[][]{%
			\label{fig-A2hatRe}%
			\includegraphics[width=\textwidth]{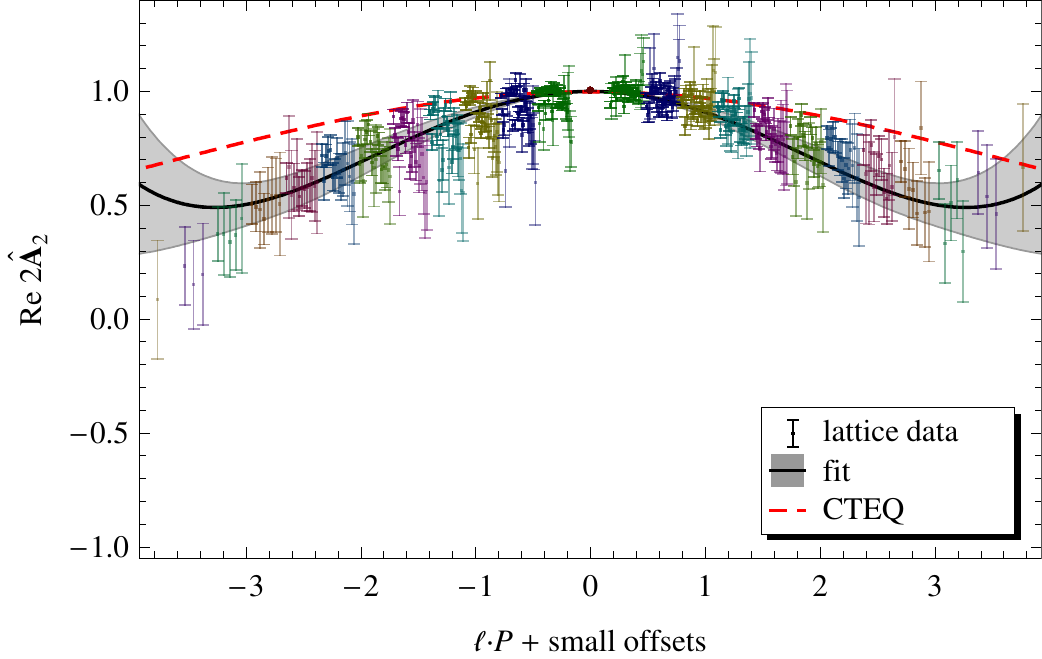}%
			}\\
		\subfloat[][]{%
			\label{fig-A2hatIm}%
			\includegraphics[width=\textwidth]{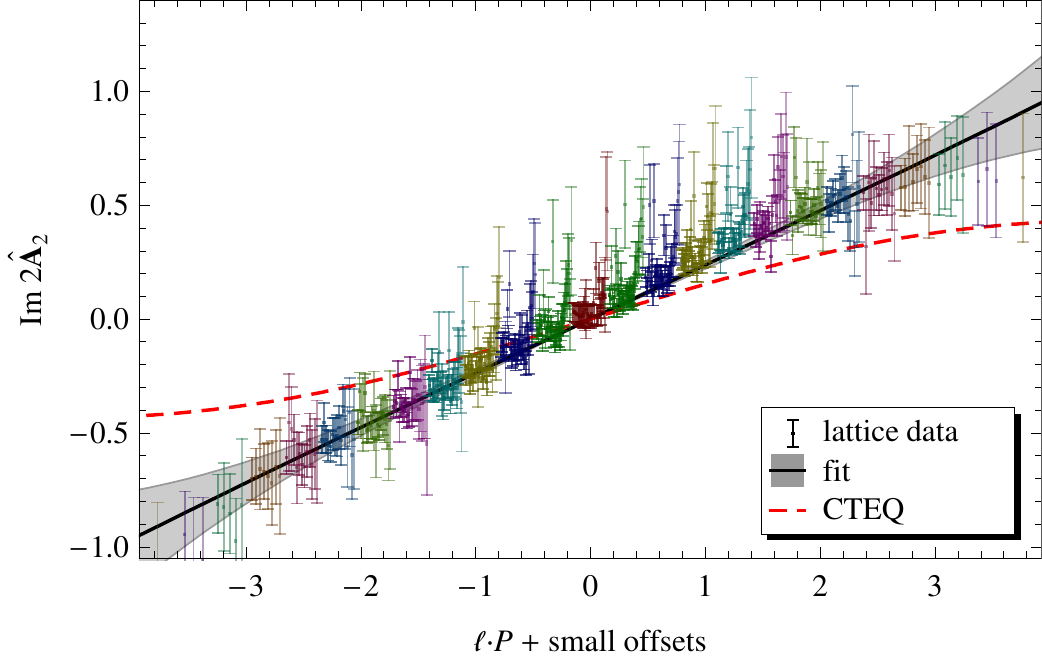}%
			} 
	\end{minipage}%
	\caption[factorization test]{%
		\subref{fig-facttest}: Test of factorization for $\myIm \tilde A_2$ at $m_\pi \approx 600 \units{MeV}$ for up minus down quarks. The statistical errors of the data points shown are correlated. No statistically significant deviation from factorization (dashed lines) is visible.\par%
		\subref{fig-A2hatRe} and \subref{fig-A2hatIm}: %
		$\AtwoldP(\elll^2, \elll\tcdot P)$ for up minus down quarks plotted with respect to $\elll\tcdot P$. At each value of $\elll\tcdot P$ (indicated by color), results for a number of values of $\elll^2$ are plotted. Small offsets have been introduced for clarity. If factorization were strongly violated, the data points at a given value of $\elll \tcdot P$ would not lie close together. The dashed curve is a Fourier transform of the CTEQ5M parton distribution function $f_1(x)$ \cite{CTEQ5} at a scale of $Q^2 = (2\units{GeV})^2$.
		The gray bands are polynomial fits to the lattice data. In \subref{fig-A2hatRe}, the fit function is of the form $1+c_2 (\elll \tcdot P)^2 + c_4 (\elll \tcdot P)^4$, and in \subref{fig-A2hatIm} it is of the form $c_1 (\elll \tcdot P) + c_3 (\elll \tcdot P)^3$.
		\label{fig-A2fact} 
		}
	\end{figure}

In phenomenological applications, it is often assumed that $f_1(x,k_\prp)$ factorizes into an $x$- and a $k_\prp$-dependent part, see, e.g., \cite{Anselmino:2005nn}. In our case, the hypothesis that $f_1^\lat$ is of the form $f_1^\lat(x,k_\prp) = \fonex(x)\ f_1^{(1)\lat}\!(k_\prp)$ translates into $\tilde A_2(\elll^2,\elll \tcdot P) = \AtwoldP(\elll\tcdot P)\ \tilde A_2(\elll^2,0)$ using the Fourier transform eq.\,(\ref{eq-f1ft}). To test this hypothesis, we introduce a scaled amplitude
\begin{equation}
	\AtwoldP(\elll^2, \elll\tcdot P)\ \equiv\ \frac{ \tilde A_2(\elll^2,\elll \tcdot P) }{ \myRe \tilde A_2(\elll^2,0) }%
	.
\end{equation}
Note that renormalization factors cancel in this ratio. If factorization holds, $\AtwoldP(\elll^2, \elll\tcdot P)$ should be $\elll^2$-independent. We plot this quantity in Fig.~\ref{fig-facttest}, selecting the imaginary part as an example. No significant $\elll^2$-dependence is visible, i.e., we confirm the factorization hypothesis within the accessible kinematic range and within our statistics.
Given that $\AtwoldP(\elll^2, \elll\tcdot P)$ is approximately $\elll^2$-independent, we can plot it with respect to $\elll\tcdot P$ as in Figs. \ref{fig-A2hatRe} and \ref{fig-A2hatIm}. It is interesting to see that the result is qualitatively similar to the Fourier transform $\int dx \exp(i\, \elll \tcdot P\, x) f_1(x)$ of a phenomenological parametrization of the PDF $f_1(x)$, such as the one provided by CTEQ5 \cite{CTEQ5}.

\section{Conclusions and Outlook}

Presently, we employ a straight Wilson line between the quark fields in the definition of TMDPDFs. We have shown first preliminary results from lattice QCD for the TMDPDFs $f_1^\text{sW}$ and $g_{1T}^\text{sW}$ as a function of transverse momentum. We find that densities of longitudinally polarized quarks in a transversely polarized proton are deformed. Moreover, we confirm that the factorization hypothesis $f_1^\lat(x,k_\prp) = \fonex(x)\ f_1^{(1)\lat}\!(k_\prp)$ is valid within the statistics of our data set.

Concerning our renormalization procedure, further investigations are in progress. Furthermore, we would like to extend our work towards non-straight Wilson lines, similar to those appearing in the definition of TMDPDFs for experimental processes such as SIDIS.

\acknowledgments

We thank Vladimir~Braun, Gunnar~Bali and Meinulf G{\"o}ckeler for very
helpful discussions and the members of the LHPC collaboration for providing propagators and technical expertise. B.~M. and Ph.~H. acknowledge support by the DFG Emmy Noether-program and the Excellence Cluster Universe at the TU M{\"u}nchen, A.~S. acknowledges support by BMBF. This work was supported in part by funds provided by the U.S. Department of Energy under grant DE-FG02-94ER40818.

\bibliography{bmusch_PhD}
\bibliographystyle{utphys} 


\end{document}